\documentclass[eqsecnum,preprint,prd,aps,nofootinbib]{revtex4}
\usepackage{amsmath,amsfonts}
\usepackage{graphics}
\newcommand{\be}{\begin{equation}}
\newcommand{\ee}{\end{equation}}
\newcommand{\ba}{\begin{eqnarray}}
\newcommand{\ea}{\end{eqnarray}}
\newcommand{\bc}{\begin{center}}
\newcommand{\ec}{\end{center}}
\def\lvec#1{\vbox{\ialign{##\crcr$\leftarrow$\crcr\noalign{
 \kern-1pt\nointerlineskip}$\hfil\displaystyle{#1}\hfil$\crcr}}}
\def\rvec#1{\vbox{\ialign{##\crcr$\rightarrow$\crcr\noalign{
 \kern-1pt\nointerlineskip}$\hfil\displaystyle{#1}\hfil$\crcr}}}

\begin{document}
\begin{center}
\bibliographystyle{article}

{\Large \textsc{Spinor Two-Point Functions and Peierls Bracket in 
de Sitter Space }}

\end{center}
\vspace{0.4cm}


\author{Giampiero Esposito$^{1}$ \thanks{
Electronic address: giampiero.esposito@na.infn.it}
Raju Roychowdhury$^{2,1}$ \thanks{
Electronic address: raju@na.infn.it}}

\affiliation{
${\ }^{1}$Istituto Nazionale di Fisica Nucleare, Sezione di Napoli,\\
Complesso Universitario di Monte S. Angelo, Via Cintia, Edificio 6, 80126
Napoli, Italy\\
${\ }^{2}$Dipartimento di Scienze Fisiche, Federico II University,\\
Complesso Universitario di Monte S. Angelo, Via Cintia, Edificio 6,
80126 Napoli, Italy}

\vspace{0.4cm}
\date{\today}

\begin{abstract}
This paper studies spinor two-point functions 
for spin-1/2 and spin-3/2 fields in maximally symmetric spaces 
such as de Sitter(dS) spacetime, by using 
intrinsic geometric objects. The Feynman, positive- and 
negative-frequency Green functions are then obtained for these cases,
from which we eventually display the supercommutator and
the Peierls bracket under such a setting in  
two-component-spinor language.
\end{abstract}

\maketitle
\bigskip
\vspace{2cm}

\section{Introduction}

The formulation of a theory of quantum gravity requires one to 
thoroughly understand the 
particle propagation in curved spacetimes. Maximally symmetric 
spaces such as de Sitter and anti-de Sitter
provide one with an interesting backdrop to study quantum field 
theory in curved spacetimes.
In this background geometry, if one needs to calculate basic 
quantities like scattering amplitudes,
one should find out the correlation function which involves the 
propagators for various particles in
this background. Thus, the problem of calculating the propagators 
has always been of much physical
interest to several authors. This hunt also assumed much significance 
after the advent of the famous Maldacena conjecture or the AdS/CFT 
correspondence \cite {maldacena,klebanov,witten} which proposes 
a duality between a quantum gravity theory on the bulk $AdS_{d+1}$ and 
a strongly coupled conformal d-dimensional gauge theory at large $N$ on 
the boundary of it. Then there is the recently proposed dS/CFT correspondence
\cite{strominger} which might shed light on quantum gravity in de Sitter space.
This conjecture, which is largely modeled on analogy with 
AdS/CFT \cite{volovich}, still lacks a clear relation 
to string theory which in turn hinders the explicit 
realization of the proposal made by Strominger. 
At the same time, a consistent formulation
of all interactions in de Sitter space 
is also tempting because of the recent observational data 
in favor of the inflationary picture. 

In field theory the Peierls bracket is a Poisson bracket which
is invariant under the full infinite-dimensional invariance group
of the action functional. Without invoking a definition of canonical 
coordinates and canonical momenta in advance, the Peierls bracket follows 
directly from the classical action, and is
made out of the advanced and retarded Green's functions. 
Hence it is necessary to build the spinor parallel 
propagator and the spinor Green function in order to write the
Peierls bracket in de Sitter and anti-de Sitter spaces 
for spin-1/2 and spin-3/2 particles.
Here we focus our calculation on de Sitter space. 

This paper consists of two parts, the first one involves the case of 
ordinary spin-1/2 particles, and the second part extends the same 
physics to spin-3/2 fields, i.e. the gravitino. In this pedagogical 
paper we first introduce the idea of the Peierls bracket in Sec. II, 
while Sec. III contains an introduction to 
maximally symmetric bitensors. 
In Sec. IV we review a few elementary properties of the spinor 
parallel propagator, in Sec. V we calculate the massive
spinor Green functions and hence the Feynman, positive- and
negative-frequency two-point functions. Then we
show how to build a Peierls bracket from this for the spin-1/2 case.
In Sec. VI we summarize all techniques developed so 
far in this paper and apply them to evaluate the gravitino  
Green functions in four-dimensional de Sitter space in two-component spinor  
language, and finally conclude our paper with the 
explicit construction of a Peierls bracket for gravitinos, 
which has not been done so far to our knowledge. 
Concluding remarks are presented in Sec. VII, while relevant details
are given in the Appendix.

\section{The Peierls bracket}

Since the Peierls bracket is not quite a familiar concept, a
brief review about it is presented here. For more details on the subject
we refer the reader to \cite{Bryce, Peierls, giampiero} 
and the references therein.

It was in the early fifties when R.E. Peierls 
\cite{Peierls} first noticed a similarity
in algebraic structure between the Poisson bracket and the Peierls 
bracket as it is called today (for theories without gauge freedom the 
Peierls bracket is indeed a Poisson bracket, whereas for gauge theories
it becomes a Poisson under restriction to the space of observables
\cite{Bryce, giampiero}), and found that this new structure
could be defined directly from the action principle without
performing a canonical decomposition into coordinates and
momenta.  His essential insight was to consider the
advanced and retarded ``effect of one quantity ($A$) on another ($B$)."
Here, $A$ and $B$ are functions on the space of histories ${\cal H}$.  
The space-of-histories formulation using the DeWitt condensed-index
notation often proves indeed very useful in
the study of the generalized Peierls algebra
and provides the opportunity to introduce in a concise way
the relevant techniques. We first define the advanced and
retarded effects of $A$ and $B$ on each other as functions
on ${\cal H}$, from which the Peierls bracket
follows. This will be straightforward by using the machinery
of \cite{Bryce} and indeed, much of what follows is
implicit in that treatment.

Once the action functional $S$ is replaced by a new action functional
$S+\epsilon A$ after the interaction with some external agent, the
small disturbances $\delta \phi^{i}$ are ruled by an inhomogeneous
differential equation (see below) which is solved after inverting a
differential operator $F_{ij}$. On denoting by $G^{\pm jk}$ the advanced
(resp. retarded) Green functions of $F_{ij}$, one can define
\begin{equation}
\delta_{A}^{\pm}B \equiv \epsilon B_{,i}G^{\pm ij}A_{,j}, \;
D_{A}B \equiv \lim_{\epsilon \to 0}{1\over \epsilon}
\delta_{A}^{-}B,
\label{(2.1)}
\end{equation}
and the Peierls bracket
\begin{equation}
(A,B) \equiv D_{A}B-D_{B}A.
\label{(2.2)}
\end{equation}
To be more precise, following \cite{Bryce}, recall that the 
undisturbed fields satisfy the equations of motion
\begin{equation}
\label{eom}
0 = S,_i(\phi^j)
\end{equation}
while the disturbed fields satisfy
\begin{equation}
0= S_{\epsilon},_i(\phi^j_{\epsilon}) =
S,_i(\phi^j_{\epsilon}) + \epsilon A,_i(\phi^j_{\epsilon}).
\end{equation}
To first order, the perturbations $\delta \phi^i$ are
therefore governed by the equation
\begin{equation}
\label{sd}
S,_{ij}(\phi^k) \delta \phi^{j} = -\epsilon \; A_{,i}(\phi^{k}),
\end{equation}
and we see that both the boundary conditions (advanced or
retarded) and any gauge fixing applies only to the
inversion of the operator $S,_{ij}(\phi^k)$ in the above
linear equation for $\delta \phi^j$ and not to the
solution of (\ref{eom}) for $\phi^i$.  In the case where
there are no gauge symmetries, $S,_{ij}$ is invertible and
has advanced and retarded Green's functions $G^{\pm jk}$
that satisfy \cite{Bryce}
\begin{equation}
\label{inverseS}
S,_{ij} G^{\pm jk} = - \delta^{k}_{i},
\end{equation}
so that the advanced and retarded solutions to the above
equations are $\delta ^{\pm}\phi^j = \epsilon G^{\pm ji}A,_i$ where
both $G^{\pm ji}$ and $A,_i$ depend on the unperturbed
solution $\phi^i$. From the definitions (2.1) and (2.2), 
the Peierls bracket is just
\begin{equation}
\label{alg}
(A,B) = A_{,i} \widetilde{G}^{ij} B_{,j},
\end{equation} 
where
\begin{equation}
\label{tildeG}
\widetilde{G}^{ij} \equiv G^{+ij} - G^{-ij}
\end{equation}
is called the {\it supercommutator function}, 
i.e. the difference of advanced and retarded Green functions.
 
For gauge fields, however, 
there exists on $\Phi$ a set of vector fields $Q_{\alpha}$ 
that leave the action $S$ invariant, i.e.
\begin{equation}
Q_{\alpha}S=0.
\label{(1.1)}
\end{equation}
If  $A$ and $B$ are two such gauge-invariant functionals:
\begin{equation}
Q_{\alpha}A=Q_{\alpha}B=0,
\label{(1.7)}
\end{equation}
then their Peierls bracket $(A,B)$ is defined as follows
\cite{DeWi03, Espo07}:
\begin{equation}
(A,B) \equiv A_{,i}{\widetilde G}^{ij}B_{,j}
=\int \int dx \; dy {\delta A \over \delta \varphi^{i}(x)}
{\widetilde G}^{ij}(x,y){\delta B \over \delta \varphi^{j}(y)},
\label{(1.8)}
\end{equation}
where the advanced and retarded Green functions used to define
the supercommutator ${\widetilde G}^{ij}$ now pertain to the
invertible gauge-field operator obtained from the gauge-fixing procedure.
Since  $A$ and $B$ are observables, Jacobi identity and gauge 
invariance hold for the Peierls bracket
(for a detailed proof of these properties see, for example, 
\cite {Bryce, giampiero}).

\section{Maximally symmetric bitensors}

More than two decades ago Allen and co-authors used 
intrinsic geometric objects to calculate correlation functions in maximally
symmetric spaces; their results, here exploited, 
were presented in a series of papers \cite{allen1,allen2}.
In this section we would like to review the elementary maximally symmetric 
bi-tensors which have been discussed previously by Allen and 
Jacobson \cite{allen1},
although more recently the calculation of spinor parallel propagator 
has been carried out in arbitrary dimension \cite{mueck}.

A maximally symmetric space is a topological manifold of dimension $n$, 
with a metric which has the maximum number of global Killing 
vector fields. This type of space looks exactly the same in every direction and 
at every point. The simplest examples are flat space and sphere, each of which 
has $\case{1}{2}n(n+1)$ independent Killing fields.
For $S^n$ these generate all rotations, and for 
$\mathbb{R}^n$ they include both rotations and translations.

Consider a maximally symmetric space of dimension $n$ with constant
scalar curvature $n(n-1)/R^2$. For the space $S^n$, the radius $R$ is
real and positive, whereas for the hyperbolic space $H^n$, $R=il$ with
$l$ positive, and in the flat case, $\mathbb{R}^n$, $R=\infty$. Consider 
further two points $x$ and $x'$, which can be connected
uniquely by a shortest geodesic. Let $\mu(x,x')$ be the proper geodesic
distance along this shortest geodesic between $x$ and $x'$. If $n^a(x,x')$
and $n^{a'}(x,x')$ are the tangents to the geodesic at $x$ and $x'$, 
the tangent vectors are then given in terms of the 
geodesic distance as follows:
\begin{equation}
\label{ndef} 
n_{a}(x,x') = \nabla_{a}\mu(x,x') \quad \text{and} \quad 
n_{a'}(x,x') = \nabla_{a'} \mu(x,x').
\end{equation}
Furthermore, on denoting
by $g^{a}_{\;b'}(x,x')$ the vector parallel propagator along the
geodesic, one can then write $n^{b'} = -g^{b'}_{\;a} n^a$. Tensors
that depend on two points, $x$ and $x'$, are 
bitensors \cite{synge}. They may carry unprimed or primed indices that 
live on the tangent space at $x$ or $x'$.

These geometric objects $n^a$, $n^{a'}$ and
$g^a_{\;b'}$ satisfy the following properties \cite{allen1}:
\begin{subequations}
\begin{align}
\label{dn}
 \nabla_a n_b &= A(g_{ab} -n_a n_b), \\
\label{dnprime}
 \nabla_a n_{b'} &= C(g_{ab'} +n_a n_{b'}), \\
\label{dg}
 \nabla_a g_{bc'} &= -(A+C) (g_{ab} n_{c'} +
 g_{ac'} n_b),
\end{align}
\end{subequations}
where $A$ and $C$ are functions of the geodesic distance 
$\mu$ and are given by \cite{allen1} 
\begin{equation}
\label{AC}
 A = \frac1R \cot \frac{\mu}R \quad \text{and} \quad 
 C = -\frac1{R\sin(\mu/R)}, 
\end{equation}
and thus they satisfy the relations 
\begin{equation}
\label{ACrel}
dA/d\mu =-C^2, \quad dC/d\mu =-AC \quad \text{and} \quad C^2-A^2
=1/R^2.
\end{equation}
Last, our convention for covariant gamma matrices is
\begin{equation}
\{\Gamma^\mu,\Gamma^\nu\} =2 I g^{\mu\nu}.
\label{(3.5)}
\end{equation}

\section{The spinor parallel propagator}

In this paper we follow the conventions for two-component
spinors, as well as all signature and curvature conventions, 
of Allen and Lutken \cite{allen2}, and hence
we use dotted and undotted spinors instead 
of the primed and unprimed ones of Penrose and Rindler  
\cite{penrose}. In our work a primed index indicates instead that
it lives in the tangent space at $x'$, while the unprimed ones live at $x$.
The fundamental object to deal with here is the bispinor $D_A^{\;A'}(x,x')$
which parallel transports a two-component spinor 
$\phi^A$ at the point $x$, along the 
geodesic to the point $x'$, yielding a new spinor $\chi^{A'}$ at $x'$, i.e.
\begin{equation}
\label{def}
\chi^{A'}=\phi^{A} \;  D_{A}^{\;A'}(x,x').
\end{equation}
Complex conjugate spinors are similarly transported by the complex conjugate
of $D_A^{\;A'}(x,x')$, which is $\overline{D}_{\dot{A}}^{\;{\dot{A}'}}(x,x')$.
A few elementary properties of $D_A^{\;A'}$ are listed below (some of them 
will be used for later calculations) \cite{allen2}:
\begin{subequations}
\begin{align}
\label{p1}
D_A^{\;A'}(x,x') = - D^{{A}'}_{\;A}(x',x),\\
\label{p2}
D_A^{\;A'}D_{{A}'}^{\;B} = \varepsilon_{A}^{\;B},\\
\label{p3}
D_{AA'}D^{AA'} = 2, \\
\label{p4}
\lim_{x\to x'} D_{A}^{\;B'} = \varepsilon_{A}^{\;B},\\
\label{p5}
g_a^{\;b'} = D_A^{\;B'}\overline{D}_{\dot{A}}^{\;{\dot{B}'}},\\
\label{p6}
D_A^{\;B'}\overline{D}_{\dot{A}}^{\;{\dot{B}'}} 
n^{A\dot{A}} = - n^{B'{\dot{B}}'}, \\
\label{p7}
n_{A\dot{C}}D^A_{\;B'} = - n_{B'{\dot{B}}'}
\overline{D}^{{\dot{B}}'}_{\;{\dot{C}}}, \\
\label{p8}
\nabla_{A\dot{A}}D^A_{\;A'} = \frac{3}{2}(A+C)n_{A\dot{A}} D^A_{\;A'}, \\
\label{p9}
D^A_{\;A'}\nabla_{A\dot{A}}n^{A'{\dot{A}}'} 
= - \frac{3}{2}C\overline{D}_{\dot{A}}^{\;{\dot{A}'}}, \\
\label{p10}
\nabla_{A\dot{A}}n^A_{\;\dot{B}} = \frac{3}{2}A\varepsilon_{\dot{A}\dot{B}}.
\end{align}
\end{subequations}
Just to recall the previously defined notations and set up the 
two-component formalism, we note from (3.1) that $n_{A\dot{A}} = 
\nabla_{A\dot{A}}\mu$ and 
$n_{A'{\dot{A}}'} = \nabla_{A'{\dot{A}}'}\mu$, where 
$\mu(x,x')$ is the geodesic separation of $x$ and $x'$. For 
completeness we should also find the covariant derivative of 
$D_A^{\;A'}$, which is formed out of the tangent 
$n_{A{\dot A}}$ to the geodesic and from $D_A^{\;A'}$ itself, i.e.
\begin{equation}
\label{defcovderarbritary}
\nabla_{A\dot{A}}D_B^{\;B'} \equiv \alpha(\mu)n_{A\dot{A}}D_B^{\;B'}
+\beta(\mu)n_{B\dot{A}}D_A^{\;B'}.
\end{equation}
Here $\alpha$ and $\beta$ are two arbitrary functions of the 
geodesic distance to be determined. 
But both of them are not independent and are 
related to each other because of the fact
that $D_B^{\;B'}$ and $n_{A\dot{A}}$, by definition, satisfy the 
following relations \cite{allen2}:
\begin{equation} 
\label{defparallelprop}
n^a\nabla_aD_B^{\;B'} = 0,
\end{equation}
\begin{equation}
\label{defn}
n_{A\dot{A}}n^{B\dot{A}} = \frac{1}{2}\delta_{A}^{\;B}.
\end{equation}
From the relations (\ref{defparallelprop}) and (\ref{defn}) it follows 
that $\beta(\mu) = -2\alpha(\mu)$. One
then determines $\beta(\mu)$ by using the Ricci identity, 
i.e. the integrability condition for spinors \cite{penrose}, and after 
all dust gets settled one obtains the final form of the covariant 
derivative of the spinor parallel propagator as
\begin{equation} 
\label{defcovder}
\nabla_{A\dot{A}}D_B^{\;B'}= (A+C)\left[
\frac{1}{2}n_{A\dot{A}}D_B^{\;B'}-n_{B\dot{A}}D_A^{\;B'}\right],
\end{equation}
where $A$ and $C$ are defined as in the previous section.

\section{The spinor Green function}

First we define a four-component Dirac spinor by 
\begin{equation}
\label{defspinor}
\psi_{\alpha}= \mbox{$ \left( \begin{array}{c}
\phi_{A} \\
\overline{\chi}_{\dot{A}}\end{array} \right) $} , 
\end{equation}
where $\phi_{A}$ and $\overline{\chi}_{\dot{A}} $ are a pair of two-component 
spinors satisfying the Dirac equation \cite{penrose}
\begin{equation}
\label{eqnphi}
\nabla_{A\dot{A}}\phi^{A} = \frac{-m}{\sqrt{2}}\overline{\chi}_{\dot{A}},
\end{equation}
\begin{equation}
\label{eqnchi}
\nabla_A^{\;\dot{A}}\overline{\chi}_{\dot{A}} = \frac{m}{\sqrt{2}}\phi_{A},
\end{equation}
$m$ being the mass of the spin-1/2 field.
We can define two basic massive two-point functions, which are 
\begin{equation}
\label{defP}
P^{A{\dot{B}}'} = \langle\phi^{A}(x)
\overline{\phi}^{{\dot{B}}'}(x')\rangle = f(\mu)D^A_{\;A'}n^{A'{\dot{B}}'},
\end{equation}
\begin{equation}
\label{defQ}
Q_{\dot{A}}^{{\;\dot{B}}'} = \langle
\overline\chi_{\dot{A}}(x)\overline{\phi}^{{\dot{B}}'}(x')
\rangle = g(\mu)\overline{D}_{\dot{A}}^{\;{\dot{B}'}}.
\end{equation}
Here we temporarily assume the spacelike separation between 
the points $x$ and $x'$ such that the field operators in 
(\ref{defP}) and (\ref{defQ}) anti-commute. On the right-hand 
side of  (\ref{defP}) and (\ref{defQ}) we have the most general 
maximally symmetric bispinor with the correct index structure. 
It is to be noted that the functions $f$ and $g$ appearing here in 
the structure, do depend only on the geodesic distance $\mu$, and other 
two-point functions like $\langle\overline\chi_{\dot{A}}\chi_{B'}\rangle$ and 
$\langle\phi^A\chi_{B'}\rangle$ are entirely determined by $f$ and $g$ only.
The equations of motion (\ref{eqnphi}) and (\ref{eqnchi}) now imply that
\begin{equation}
\label{nablaP}
\nabla_{A\dot{A}}P^{A{\dot{B}}'} 
= \frac{-m}{\sqrt{2}}Q_{\dot{A}}^{{\;\dot{B}}'},
\end{equation}
\begin{equation}
\label{nablaQ}
\nabla_A^{\;\dot{A}}Q_{\dot{A}}^{{\;\dot{B}}'} 
= \frac{m}{\sqrt{2}}P_{A}^{{\;\dot{B}}'}.
\end{equation}
If now we insert equations (\ref{defP}) and (\ref{defQ}) into 
equations (\ref{nablaP}) and (\ref{nablaQ}) we obtain, after a 
little gymnastics with two-spinor calculus, two coupled equations 
for the coefficients $f(\mu)$ and $g(\mu)$ as follows:
\begin{equation}
\label{eqf}
f'+\frac{3}{2}(A-C)f+\sqrt{2}mg = 0,
\end{equation}
\begin{equation}
\label{eqg}
g'+\frac{3}{2}(A+C)g-\frac{m}{\sqrt{2}}f = 0,
\end{equation}
where the prime stands for derivative with respect to $\mu$.
On differentiating (\ref{eqf}) with respect to $\mu$ once and then 
using (\ref{ACrel}) and (\ref{eqg}) successively one finds a 
second-order equation for $f$: 
\begin{equation}
\label{forder2}
f''(\mu)+3Af'(\mu)+\left[m^2-\frac{9}{4}R^{-2}
+\frac{3}{2}C(A-C)\right]f(\mu) = 0.
\end{equation}
Now to solve for $f(\mu)$ and $g(\mu)$, one makes a change of variable 
\begin{equation}
\label{cov}
Z \equiv \cos^2\left(\frac{\mu}{2R}\right)
\end{equation}
to write (\ref{forder2}) as \cite{allen2} 
\begin{equation}
\label{newfequation}
Z(1-Z)\frac{d^2}{dZ^2}f(Z)+2(1-2Z)\frac{d}{dZ}f(Z)
+\left[m^2R^2-\frac{9}{4}-\frac{3}{4(1-Z)}\right]f(Z) = 0.
\end{equation}
On making further a redefinition
\begin{equation}
\label{defw} 
w(Z) \equiv [R^2(1-Z)]^{-1/2}f(Z),
\end{equation}
one rewrites (\ref{newfequation}) as a hypergeometric equation 
in the variable $w$, i.e.
\begin{equation}
\label{hypergeometric}
H(a,b,c;Z)w(Z) = 0,
\end{equation}
where $H(a,b,c)$ is the hypergeometric operator
\begin{equation}
\label{hypergeoop}
H(a,b,c;Z) = Z(1-Z)\frac{d^2}{dZ^2}+[c-(a+b+1)Z]\frac{d}{dZ}-ab.
\end{equation}
Following our source,
the factor $R^2$ is included in the definition (\ref{defw}) of $w$ 
to ensure that the standard branch cut of the square root function 
lies along the timelike separations $\mu^2 >0$.
The parameters $a,b,c$ here take the values 
\begin{subequations}
\begin{align}
a = 2+\sqrt{m^2R^2},\\
b = 2-\sqrt{m^2R^2},\\
c = 2.
\end{align}
\end{subequations}
In the same way it can be shown that if we let $w(Z) = 
[R^2(Z)]^{-1/2}g(Z)$, then $w$ satisfies a hypergeometric equation 
with parameters $a$,$b$ and $c+1$. Now one has to specify the boundary 
conditions to uniquely specify a solution to the hypergeometric equation.
The correct solution to (\ref{hypergeometric}) in de Sitter space 
$R^2 < 0$ is obtained (following \cite{allen1}) by demanding that it is 
only singular when $\mu=0$, that is $Z=1$, and not when $\mu=\pi R$, 
that is $Z=0$. Two independent solutions of the hypergeometric equations 
\cite{abra,erdelyi} are therefore $F(a,b;c;Z)$ and $F(a,b;c+1;Z)$, 
and this yields the following solutions: 
\begin{equation}
\label{soln1}
f_{DS} = N_{DS}(1-Z)^{1/2}F(a,b;c;Z),
\end{equation}
\begin{equation}
\label{soln2}
g_{DS}= -iN_{DS}2^{-3/2}m|R|Z^{1/2}F(a,b;c+1;Z).
\end{equation}
The short distance behavior $\mu \to 0$ can now be used to fix the 
as yet undetermined constant 
$N_{DS}$. The flat-space limit as $\mu \to 0$ is 
\begin{equation}
\label{flatspacelim}
f \sim \frac{-i}{\sqrt{2}} \frac{1}{\pi^2}(-\mu^2)^{-3/2}.
\end{equation}
Thus, from (\ref{soln1}) it follows that 
\begin{equation}
\label{NDS}
N_{DS} = \frac{f_{DS}}{(1-Z)^{1/2}F(a,b;c;Z)}.
\end{equation}
Furthermore, near $Z = 1$ we have 
\begin{equation}
\label{Feqn}
F(a,b;c;Z) \sim \frac{\Gamma(c)\Gamma(a+b-c)}
{\Gamma(a)\Gamma(b)}(1-Z)^{c-a-b},
\end{equation}
and $(1-Z) = (\mu/2R)^{2}$, hence one finds that 
\begin{equation}
\label{NDSexplicit}
N_{DS} = \frac{-i}{\sqrt{2}} \frac{(-\mu^2)^{-3/2}}
{\pi^2}\frac{\Gamma(a)\Gamma(b)}{\Gamma(c)\Gamma(a+b-c)}
(1-Z)^{a+b-c-\frac{1}{2}}
=  \frac{-i}{\sqrt{2}} \frac{(-\mu^2)^{-3/2}}{\pi^2}
\Gamma(a)\Gamma(b)\frac{\mu^{3}}{8R^3},
\end{equation}
where we have used the fact that $\Gamma(a+b-c)=\Gamma(2)=1$ and 
similarly $\Gamma(c)=1$. On
using the values of $a$ and $b$ and putting them together in the 
expression (\ref{NDSexplicit}) one gets
\begin{equation}
N_{DS} = \frac{-i\Gamma(2+\sqrt{m^{2}R^{2}}) 
\Gamma(2-\sqrt{m^{2}R^{2}})}{8\sqrt{2}\pi^{2}|R|^{3}}.
\end{equation}
Furthermore, from the relations $\Gamma(z+1)=z\Gamma(z)$ and 
$\Gamma(1+i|mR|)\Gamma(1-i|mR|) = 
\frac{\pi|Rm|}{\sinh(\pi |Rm|)}$ one can rewrite the final answer 
for the constant $N_{DS}$
\begin{equation}
\label{NDSfinalform}
N_{DS} = \frac{-i|Rm|(1-m^{2}R^{2})}{8\sqrt{2}\pi|R|^{3}\sinh\pi|Rm|}.
\end{equation}
Once we determine $N_{DS}$, the Feynman Green function is obtained 
by evaluating $f_{DS}(Z)$ and 
$g_{DS}(Z)$ above the branch cut from $Z=1$ to $\infty$, 
i.e. by taking $f_{DS}(Z+i0)$ and $g_{DS}(Z+i0)$.
This is what happens in the de Sitter case. To conclude we have the 
following two-point functions:
\begin{equation}
\label{2ptfn1}
P^{A{\dot{B}}'}_{(F)} = \lim_{\epsilon \to 0^{+}}f_{DS}
(Z+i\epsilon)D^A_{\;A'}n^{A'{\dot{B}}'},
\end{equation}
\begin{equation}
\label{2ptfn2}
Q^{\dot{A}{\dot{B}}'}_{(F)} = \lim_{\epsilon \to 0^{+}}
g_{DS}(Z+i\epsilon)\overline{D}^{\dot{A}{\dot{B}}'},
\end{equation}
where $(F)$ stands for the Feynman Green functions.

It is now helpful to recall 
the discussion of various types of Green functions 
depending on the contours in the complex $p^{0}$-plane for the integral 
representation of the Green function for the simpler case of scalar fields,
following \cite{Bryce}. From various contours the following relations 
among different Green's functions can be easily established:
\begin{subequations}
\begin{align}
\label{feynman}
G_{F} = G^{-}+G^{(-)} =  G^{+}-G^{(+)},\\
\label{positive}
G^{(+)}(x,x') = -\theta(x,x')G_{F}(x,x')+\theta(x',x)G_{F}^{*}(x,x'),\\
\label{negative}
G^{(-)}(x,x') = \theta(x',x)G_{F}(x,x')-\theta(x,x')G_{F}^{*}(x,x'),\\
\label{supercom}
\widetilde{G} = (G^{+}-G^{-}) = (G^{(+)}+G^{(-)})
=-2 \Bigr(\theta(x,x')-\theta(x',x)\Bigr){\rm Re}G_{F}.
\end{align}
\end{subequations}
With a standard notation, $G^{+}$ and $G^{-}$ are the 
advanced and retarded functions respectively, and their difference 
$\widetilde{G}$ is the  {\it supercommutator function}. 
$G_{F}$ is the Feynman Green function and $G_{F}^{*}$ is its complex 
conjugate. $G^{(+)}$ and $G^{(-)}$ are the positive- and 
negative-frequency parts, respectively. 
The $\theta(x,x')$ used above in the definition of 
advanced and retarded functions is the step function.

Now our approach to arrive at the Peierls bracket in the de Sitter case 
will be as follows: once we determine using (\ref{2ptfn1}) and 
(\ref{2ptfn2}) the Feynman Green function, instead of using the advanced 
and retarded functions, we can use (\ref{positive}) 
and (\ref{negative}) respectively to get $G^{(+)}$ and $G^{(-)}$, and 
then add them to get the {\it supercommutator function} $\widetilde{G}$. 
Then we use (2.11) to build the Peierls bracket $(\;,\;)_{P}$ 
which, in terms of the spinor fields 
\begin{equation}
\psi_{\alpha}=
\begin{pmatrix} 
\phi_{A} \\
{\overline \chi}_{{\dot A}}
\end{pmatrix}, \;
\chi_{\beta'}=
\begin{pmatrix}
\rho_{B'} \\
{\overline \sigma}_{{\dot B}'}
\end{pmatrix}
\label{(5.28)}
\end{equation}
reads as
\begin{equation}
(\psi,\chi)_{P} \equiv \int \int P(x,x') \sqrt{-g(x)} \sqrt{-g(x')}
d^{4}x \; d^{4}x',
\label{(5.29)}
\end{equation}
where
\begin{equation}
P(x,x') \equiv -2 \Bigr(\theta(x,x')-\theta(x',x)\Bigr)
\psi_{\nabla} ({\rm Re}G_{F})\chi_{\nabla},
\label{(5.30)}
\end{equation}
having set
\begin{equation}
\psi_{\nabla} ({\rm Re}G_{F})\chi_{\nabla} \equiv
\Bigr(\nabla_{A{\dot A}}{\overline \phi}^{{\dot A}}\Bigr) 
{\rm Re}P_{(F)}^{A{\dot B}'} 
\Bigr(\nabla_{B'{\dot B}'}\sigma^{B'}\Bigr)
+\Bigr(\nabla_{A{\dot A}}\chi^{A}\Bigr) 
{\rm Re}Q_{(F)}^{{\dot A}{\dot B}'}
\Bigr(\nabla_{B'{\dot B}'}\rho^{B'}\Bigr).
\label{(5.31)}
\end{equation}

\section{Massive spin-3/2 propagator}

In this section we consider the propagator of the massive 
spin-3/2 field. Let us denote the gravitino field by 
$\Psi^{\alpha}_{\lambda} (x)$. In a maximally symmetric state 
$|\, s >$ the propagator is 
\begin{equation}
\label{correlator}
S^{\alpha \beta^{\prime}}_{\lambda \nu^{\prime}} 
(x, x^{\prime}) = <s\,| \Psi^{\alpha}_{\lambda} (x) 
\Psi^{\beta^{\prime}}_{\nu^{\prime}} (x^{\prime}) |\,s> .
\end{equation}
The field equations imply that $S$ satisfies 
\begin{equation}
\label{EoM} 
(\Gamma^{\mu \rho \lambda} D_{\rho} - m \, 
\Gamma^{\mu \lambda})^{\alpha}{}_{\gamma} 
S_{\lambda \nu^{\prime}}{}^{\gamma}{}_{\beta^{\prime}} =
\frac{\delta (x-x^{\prime})}{\sqrt{-g}} 
g^{\mu}{}_{\nu^{\prime}} \, \delta^{\alpha}{}_{\beta^{\prime}}. 
\end{equation}

\subsection{The ten gravitino invariants in two-component-spinor language}

It is very convenient to decompose the gravitino propagator in terms 
of independent structures
constructed out of $n_\mu, n_{\nu'}, g_{\mu\nu'}$ and 
$\Lambda^\alpha_{~\beta'}$.
Thus, the propagator can be written in geometric way following Anguelova 
et al. \cite{anguelova} (see also \cite{Basu}):
\begin{eqnarray}
\label{ansatz}
S_{\lambda \nu^{\prime}}{}^{\alpha}{}_{\beta^{\prime}} 
&=& \alpha (\mu) \, g_{\lambda \nu^{\prime}} 
\Lambda^{\alpha}{}_{\beta^{\prime}} + 
\beta (\mu) \, n_{\lambda} n_{\nu^{\prime}} 
\Lambda^{\alpha}{}_{\beta^{\prime}} + 
\gamma (\mu) \, g_{\lambda \nu^{\prime}} (n_{\sigma} 
\Gamma^{\sigma} \Lambda)^{\alpha}{}_{\beta^{\prime}} \nonumber \\ 
&& + \delta (\mu) \, n_{\lambda} n_{\nu^{\prime}} (n_{\sigma} 
\Gamma^{\sigma} \Lambda)^{\alpha}{}_{\beta^{\prime}} + 
\varepsilon (\mu) \, n_{\lambda} (\Gamma_{\nu^{\prime}} 
\Lambda)^{\alpha}{}_{\beta^{\prime}} + 
\theta (\mu) \, n_{\nu^{\prime}} (\Gamma_{\lambda} 
\Lambda)^{\alpha}{}_{\beta^{\prime}} 
\nonumber \\ 
&& + \tau (\mu) \, n_{\lambda} (n_{\sigma} \Gamma^{\sigma} 
\Gamma_{\nu^{\prime}} \Lambda)^{\alpha}{}_{\beta^{\prime}} 
+ \omega(\mu)\, n_{\nu^{\prime}} (n_{\sigma} \Gamma^{\sigma} 
\Gamma_{\lambda} \Lambda)^{\alpha}{}_{\beta^{\prime}} 
\nonumber \\ 
&& + \pi (\mu) \, (\Gamma_{\lambda} 
\Gamma_{\nu^{\prime}} \Lambda)^{\alpha}{}_{\beta^{\prime}}  
+ \kappa (\mu)\, (n_{\sigma} \Gamma^{\sigma} 
\Gamma_{\lambda} \Gamma_{\nu^{\prime}} \Lambda)^{\alpha}{}_{\beta^{\prime}}. 
\end{eqnarray}
From here on we will be trying to re-write each of the building 
blocks of the invariant structure in two-spinor language and finally 
construct the full invariant propagator in two-spinor form in harmony 
with the spin-1/2 propagator previously discussed.  
Following Allen and Lutken \cite{allen2} we can write the 
gamma matrix in two-spinor language as follows (Penrose and Rindler,
on page 221 of \cite{penrose}, do not have the $-i$ factor since
their $\gamma$-matrices satisfy the anti-commutation relation 
$\gamma_{a}\gamma_{b}+\gamma_{b}\gamma_{a}=-2I g_{ab}$, unlike the
sign convention in our Eq. (3.5))
\begin{equation}
\label{gamma}
(\gamma_{p})_{\alpha}^{\;\beta} = -i{\sqrt{2}} 
\begin{pmatrix}
0 & \varepsilon_{PA} \varepsilon_{\dot{P}}^{\;\dot{B}}  \\
\varepsilon_{\dot{P}\dot{A}}\varepsilon_{P}^{\;B}& 0 
\end{pmatrix}, 
\end{equation}
where $\varepsilon_{BC}$ is the curved epsilon symbol 
which raises and lowers indices within each spin-space,
is skew-symmetric and 
encodes information on the curved spacetime metric.
In the case of flat Minkowski spacetime it reduces to the well known form
\begin{equation}
\epsilon_{\Large\textsc {BC}} = \epsilon_{{\Large\textsc {B}}A} 
\epsilon_{{\Large\textsc {C}}}^{\;A}= 
\begin{pmatrix}
0 & 1 \\
-1 & 0 
\end{pmatrix}.
\label{(6.5)}
\end{equation}
From the rules of two-spinor calculus and from 
the treatment of Allen and Lutken \cite{allen2} we already know
the following correspondences:
\begin{subequations}
\begin{align}
\label{n}
n_{\alpha} \longrightarrow n_{A\dot{A}},\\
\label{n'}
n_{\beta'} \longrightarrow n_{B'{\dot{B}}'},\\
\label{g}
g_{\alpha\beta'} \longrightarrow D_{AB'}
\overline{D}_{\dot{A}{\dot{B}}'} .
\end{align}
\end{subequations}
We also know the form of the spinor parallel propagator, 
which acts according to
\begin{subequations}
\begin{align}
\label{def1}
\chi^{A'}(x') = D_{A}^{\;A'}(x,x')\phi^{A}(x),\\
\label{def2}
\chi^{A}(x) = D^{A}_{\;B'}(x,x')\phi^{B'}(x'),\\
\label{def3}
\phi^{\dot{A}}(x) = \overline{D}^{\dot{A}}_{\;\dot{B'}}(x,x')
\chi^{\dot{B'}}(x') .
\end{align}
\end{subequations}
To translate the above set of equations, e.g. 
$\phi^{\alpha} = \Lambda^\alpha_{~\beta'}\phi^{\beta'}$, 
into two-spinor language, 
both left- and right-hand sides should involve a 
\mbox{$ \left( \begin{array}{c}
\chi \nonumber\\
\phi\end{array} \right) $} column vector, with upstairs indices 
at $x$ and $x'$ respectively. Written in matrix notation we can 
combine them into one reading, i.e.
\begin{equation}
\begin{pmatrix}
\chi^{A} \\
\phi^{\dot{A}} 
\end{pmatrix} =  
\begin{pmatrix}
0 & D^{A}_{\;B'} \\
\overline{D}^{\dot{A}}_{\;\dot{B'}} & 0 
\end{pmatrix} 
\begin{pmatrix}
\chi^{\dot{B}'} \\
\phi^{B'} 
\end{pmatrix} .
\label{(6.8)}
\end{equation}
Therefore from now on we redefine $\Lambda^\rho_{~\beta'}$ 
to be a $(2 $x$ 2)$ matrix, expressed in two-spinor language as 
$\Lambda^{R\dot{R}}_{\;B'{\dot{B}}'}$ and satisfying 
the correspondence rule
\begin{equation}
\label{lambda}
\Lambda^\rho_{~\beta'} \longrightarrow 
\begin{pmatrix}
0 & D^{R}_{\;B'} \\
\overline{D}^{\dot{R}}_{\;\dot{B'}} & 0 
\end{pmatrix} .
\end{equation}
Similarly, we go on translating each of the bits of the invariant 
structure into two-spinor notation. The next one is 
$(\Gamma^{\sigma} \Lambda)^{\alpha}{}_{\beta^{\prime}}$.
We note the following translation:
\begin{equation}
\label{gammalambda}
(\Gamma^{\sigma} \Lambda)^{\alpha}{}_{\beta^{\prime}} 
= (\Gamma^{\sigma})^{\alpha}{}_{\rho}\Lambda^{\rho}{}_{\beta^{\prime}} 
\longrightarrow (\Gamma^{S\dot{S}})^{\alpha}{}_{\rho}
\Lambda^{\rho}{}_{\beta^{\prime}} .
\end{equation}
Therefore, on using the two-spinor form (6.4) of the gamma matrix and the 
two-spinor version (6.9) of the spinor parallel propagator we get
\begin{equation}
\label{gammalambdamatrix}
(\Gamma^{\sigma} \Lambda)^{\alpha}{}_{\beta^{\prime}} 
\longrightarrow -i{\sqrt{2}} 
\begin{pmatrix}
0 & \varepsilon^{S}_{\;R} \varepsilon^{\dot{S}\dot{A}}  \\
\varepsilon^{\dot{S}}_{\;{\dot{R}}}\varepsilon^{SA} & 0 
\end{pmatrix} 
\begin{pmatrix}
0 & D^{R}_{\;B'} \\
\overline{D}^{\dot{R}}_{\;\dot{B'}} & 0 
\end{pmatrix} 
= -i {\sqrt{2}} 
\begin{pmatrix}
\varepsilon^{S}_{\;R} \varepsilon^{\dot{S}\dot{A}}
\overline{D}^{\dot{R}}_{\;\dot{B'}} & 0  \\
0 & \varepsilon^{\dot{S}}_{\;{\dot{R}}}\varepsilon^{SA}D^{R}_{\;B'} 
\end{pmatrix}.
\end{equation}
Similarly, we find
 \begin{equation}
\label{ngammalambda}
(n_{\sigma}\Gamma^{\sigma} \Lambda)^{\alpha}{}_{\beta^{\prime}} 
\longrightarrow n_{S\dot{S}}(\Gamma^{S\dot{S}})^{\alpha}{}_{\rho}
\Lambda^{\rho}{}_{\beta^{\prime}} 
= -i n_{S\dot{S}}{\sqrt{2}}  
\begin{pmatrix}
\varepsilon^{S}_{\;R} \varepsilon^{\dot{S}\dot{A}}
\overline{D}^{\dot{R}}_{\;\dot{B'}} & 0  \\
0 & \varepsilon^{\dot{S}}_{\;{\dot{R}}}\varepsilon^{SA}D^{R}_{\;B'} 
\end{pmatrix}.
\end{equation}
Now we use the antisymmetry property of the  
epsilon symbol, i.e. $\varepsilon^{AB} = - \varepsilon^{BA}$, and the 
rules for raising and lowering spinor indices, i.e. 
$\varepsilon^{AB}\phi_{B} = \phi^{A}$, 
$\phi^{A}\varepsilon_{AB} = \phi_{B}$,    
to write $(n_{\sigma}\Gamma^{\sigma} \Lambda)^{\alpha}{}_{\beta^{\prime}}$ 
in matrix form as 
\begin{equation}
\label{ngammalambdamatrix}
(n_{\sigma}\Gamma^{\sigma} \Lambda)^{\alpha}{}_{\beta^{\prime}} 
\longrightarrow -i {\sqrt{2}} 
\begin{pmatrix}
n_{R}^{\;\dot{A}}\overline{D}^{\dot{R}}_{\;\dot{B'}} & 0  \\
0 & n^{A}_{\;\dot{R}}D^{R}_{\;B'} 
\end{pmatrix}.
\end{equation}

Now let us start writing the invariants in two-spinor language.
The first invariant structure (see (6.3) from now on) is
\begin{equation}
\label{inv1}
g_{\lambda \nu^{\prime}} \Lambda^{\alpha}{}_{\beta^{\prime}} 
\longrightarrow D_{LN'}\overline{D}_{\dot{L}{\dot{N}}'} 
\begin{pmatrix}
0 & D^{A}_{\;B'} \\
\overline{D}^{\dot{A}}_{\;\dot{B'}} & 0 
\end{pmatrix}.
\end{equation}
The second one is 
\begin{equation}
\label{inv2}
n_{\lambda} n_{\nu^{\prime}} \Lambda^{\alpha}{}_{\beta^{\prime}}  
\longrightarrow n_{L\dot{L}}n_{N'{\dot{N}}'}
\begin{pmatrix}
0 & D^{A}_{\;B'} \\
\overline{D}^{\dot{A}}_{\;\dot{B'}} & 0 
\end{pmatrix}.
\end{equation}
Then the third reads as
\begin{equation}
\label{inv3}
g_{\lambda \nu^{\prime}} (n_{\sigma} \Gamma^{\sigma} 
\Lambda)^{\alpha}{}_{\beta^{\prime}} 
\longrightarrow  -i D_{LN'}
\overline{D}_{\dot{L}{\dot{N}}'}{\sqrt{2}} 
\begin{pmatrix}
n_{R}^{\;\dot{A}}\overline{D}^{\dot{R}}_{\;\dot{B'}} & 0  \\
0 & n^{A}_{\;\dot{R}}D^{R}_{\;B'} 
\end{pmatrix}.
\end{equation}
Next is the fourth invariant, i.e.
\begin{equation}
\label{inv4}
n_{\lambda} n_{\nu^{\prime}} (n_{\sigma} 
\Gamma^{\sigma} \Lambda)^{\alpha}{}_{\beta^{\prime}} \longrightarrow  
-i n_{L\dot{L}}n_{N'{\dot{N}}'}{\sqrt{2}} 
\begin{pmatrix}
n_{R}^{\;\dot{A}}\overline{D}^{\dot{R}}_{\;\dot{B'}} & 0  \\
0 & n^{A}_{\;\dot{R}}D^{R}_{\;B'} 
\end{pmatrix}.
\end{equation}
The subsequent invariant structure involves $(\Gamma_{\nu^{\prime}} 
\Lambda)^{\alpha}{}_{\beta^{\prime}}$ and we know that 
\begin{equation}
\label{exception}
(\Gamma_{\nu^{\prime}} \Lambda)^{\alpha}{}_{\beta^{\prime}} 
\longrightarrow (\Gamma_{N'{\dot{N}}'})^{\alpha}{}_{\rho}
\Lambda^{\rho}{}_{\beta^{\prime}} .
\end{equation}
Now from our previous discussion in this section we already have 
\begin{equation}
\label{gammaprimematrix}
(\Gamma_{N'{\dot{N}}'})^{\alpha'}{}_{\rho'} \longrightarrow 
-i {\sqrt{2}} 
\begin{pmatrix}
0 & \varepsilon_{N'R'} \varepsilon_{{\dot{N}}'}^{\;{\dot{A}}'} \\
\varepsilon_{{\dot{N}}'{\dot{R}}'}\varepsilon_{N'}^{\;A'}& 0 
\end{pmatrix}.
\end{equation}
The problem is that what is well defined is either 
$(\Gamma_{\nu})^{\alpha}{}_{\rho}$ or 
$(\Gamma_{\nu^{\prime}})^{\alpha^{\prime}}{}_{\rho^{\prime}}$, where
everything is evaluated at the same spacetime point 
(either $x$ or $x'$). However, here the relevant invariant 
consists of a mixed structure of the kind 
$(\Gamma_{\nu^{\prime}})^{\alpha}{}_{\rho}$, and, to understand what is 
meant by it, we should use the parallel displacement bi-vector.
Eventually, with the help of some careful thought we can write 
\begin{equation}
\label{gamma'matrix}
(\Gamma_{\nu^{\prime}})^{\alpha}{}_{\rho} \longrightarrow 
(\Gamma_{N'{\dot{N}}'})^{\alpha'}{}_{\rho'}g_{\alpha^{\prime}}{}^{\alpha}
g^{\rho^{\prime}}{}_{\rho}  = -i {\sqrt{2}} 
\begin{pmatrix}
0 & \varepsilon_{N'R'} \varepsilon_{{\dot{N}}'}^{\;{\dot{A}}'} \\
\varepsilon_{{\dot{N}}'{\dot{R}}'}\varepsilon_{N'}^{\;A'}& 0 
\end{pmatrix} 
g_{A'{\dot{A}}'}{}^{A\dot{A}}g^{R'{\dot{R}}'}{}_{R\dot{R}} .
\end{equation}
Recalling the fact that $g_{A'{\dot{A}}'}{}^{A\dot{A}} = 
D^{A}_{\;A'}\overline{D}^{\dot{A}}_{\;\dot{A'}}$ and 
$g^{R'{\dot{R}}'}{}_{R\dot{R}} = D_{R}^{\;R'}
\overline{D}_{\dot{R}}^{\;\dot{R'}}$ we can write the final form of the 
matrix $(\Gamma_{\nu^{\prime}})^{\alpha}{}_{\rho}$ as follows:
\begin{equation}
\label{finalgammaprime}
(\Gamma_{\nu^{\prime}})^{\alpha}{}_{\rho} \longrightarrow 
-i {\sqrt{2}}\mbox {$ \left( 
\begin{array}{cc}
0 & -D_{RN'}\overline{D}_{\dot{R}}^{\;\dot{R'}}
D^{A}_{\;A'}\overline{D}^{{\dot{A}}}_{\;{\dot{N}}'} \\
-D_{R}^{\;R'}\overline{D}_{\dot{R}{\dot{N}}'}D^{A}_{\;N'}
\overline{D}^{\dot{A}}_{\;\dot{A'}} & 0 
\end{array} 
\right) $}.
\end{equation}
Now we can build the fifth invariant quite easily as shown here,
\begin{eqnarray}
\label{ex5}
\; & \; & 
n_{\lambda} (\Gamma_{\nu^{\prime}} \Lambda)^{\alpha}{}_{\beta^{\prime}} 
\longrightarrow -i n_{L\dot{L}}{\sqrt{2}}\mbox {$ \left( 
\begin{array}{cc}
0 & -D_{RN'}\overline{D}_{\dot{R}}^{\;\dot{R'}}D^{A}_{\;A'}
\overline{D}^{{\dot{A}}}_{\;{\dot{N}}'} \\
-D_{R}^{\;R'}\overline{D}_{\dot{R}{\dot{N}}'}D^{A}_{\;N'}
\overline{D}^{\dot{A}}_{\;\dot{A'}} & 0 \end{array} \right) $}
\mbox{$ \left( \begin{array}{cc}
0 & D^{R}_{\;B'} \\
\overline{D}^{\dot{R}}_{\;\dot{B'}} & 0 \end{array} \right) $}
\nonumber \\ 
& = & 
-i n_{L\dot{L}}{\sqrt{2}}\mbox {$ \left( \begin{array}{cc}
-D_{RN'}\overline{D}_{\dot{R}}^{\;\dot{R'}}
\overline{D}^{\dot{R}}_{\;\dot{B'}}D^{A}_{\;A'}
\overline{D}^{{\dot{A}}}_{\;{\dot{N}}'} & 0 \\
0 & -D_{R}^{\;R'}D^{R}_{\;B'}\overline{D}_{\dot{R}{\dot{N}}'}D^{A}_{\;N'}
\overline{D}^{\dot{A}}_{\;\dot{A'}} \end{array} \right) $} .
\end{eqnarray}
The sixth invariant is constructed as follows:
\begin{eqnarray}
\label{constructinv6}
\; & \; &
n_{\nu^{\prime}} (\Gamma_{\lambda} \Lambda)^{\alpha}{}_{\beta^{\prime}} 
\longrightarrow n_{N'{\dot{N}}'}(\Gamma_{\lambda})^{\alpha}{}_{\rho}
\Lambda^\rho_{~\beta'} \longrightarrow 
-i n_{N'{\dot{N}}'}{\sqrt{2}} 
\mbox {$ \left( 
\begin{array}{cc}
0 & \varepsilon_{LR} \varepsilon_{\dot{L}}^{\;\dot{A}}  \\
\varepsilon_{\dot{L}\dot{R}}\varepsilon_{L}^{\;A}& 0 \end{array} \right) $} 
\mbox{$ \left( \begin{array}{cc}
0 & D^{R}_{\;B'} \\
\overline{D}^{\dot{R}}_{\;\dot{B'}} & 0 \end{array} \right) $}
\nonumber \\
& = & 
-i n_{N'{\dot{N}}'}{\sqrt{2}}\mbox {$ \left( \begin{array}{cc}
\varepsilon_{LR} \varepsilon_{\dot{L}}^{\;\dot{A}}
\overline{D}^{\dot{R}}_{\;\dot{B'}} & 0  \\
0 & \varepsilon_{\dot{L}\dot{R}}\varepsilon_{L}^{\;A}D^{R}_{\;B'} 
\end{array} \right) $}.
\end{eqnarray}

Now we start building the last four invariants step by step. 
First we express the seventh invariant 
$n_{\lambda} (n_{\sigma} \Gamma^{\sigma} \Gamma_{\nu^{\prime}} 
\Lambda)^{\alpha}{}_{\beta^{\prime}}$ in two-spinor language.
We note that
\begin{eqnarray}
\label{constructinv7}
&&n_{\lambda} (n_{\sigma} \Gamma^{\sigma} 
\Gamma_{\nu^{\prime}} \Lambda)^{\alpha}{}_{\beta^{\prime}} 
= n_{\lambda}n_{\sigma}(\Gamma^{\sigma})^{\alpha}{}_{\rho}
(\Gamma_{\nu^{\prime}})^{\rho'}{}_{\tau'}g_{\rho^{\prime}}{}^{\rho}
g^{\tau^{\prime}}{}_{\tau}\Lambda^{\tau}{}_{\beta^{\prime}} 
\nonumber\\
& \longrightarrow & -2 n_{L\dot{L}}n_{S\dot{S}} \mbox {$ \left( 
\begin{array}{cc}
0 & \varepsilon^{S}_{\;R} \varepsilon^{\dot{S}\dot{A}}  \\
\varepsilon^{\dot{S}}_{\;{\dot{R}}}\varepsilon^{SA} & 0 
\end{array} 
\right) $}\mbox {$ \left( 
\begin{array}{cc}
0 & \varepsilon_{N'T'} \varepsilon_{{\dot{N}}'}^{\;{\dot{R}}'}  \\
\varepsilon_{{\dot{N}}'{\dot{T}}'}\varepsilon_{N'}^{\;R'}& 0 
\end{array} \right) $}\nonumber\\
&\times& D^{R}_{\;R'}\overline{D}^{\dot{R}}_{\;{\dot{R}}'} D_{T}^{\;T'}
\overline{D}_{\dot{T}}^{\;{\dot{T}}'}\mbox{$ \left( \begin{array}{cc}
0 & D^{T}_{\;B'} \\
\overline{D}^{\dot{T}}_{\;\dot{B'}} & 0 \end{array} \right) $} .
\end{eqnarray}
After a little bit of algebra we arrive at the seventh invariant, i.e.
\begin{equation}
\label{inv7}
n_{\lambda} (n_{\sigma} \Gamma^{\sigma} \Gamma_{\nu^{\prime}} 
\Lambda)^{\alpha}{}_{\beta^{\prime}} \longrightarrow -2 
n_{L\dot{L}}\mbox {$ \left( \begin{array}{cc}
0 & - n_{R}^{\;\dot{A}}D^{R}_{\;N'}\overline{D}^{\dot{R}}_{\;\dot{R'}}
D_{T}{}^{T'}\overline{D}_{\dot{T}{\dot{N}}'}D^{T}_{\;B'} \\
- n^{A}_{\dot{R}}\overline{D}^{\dot{R}}_{\;{\dot{N}}'}
D^{R}_{\;R'}\overline{D}_{\dot{T}}^{\;\dot{T'}}D_{TN'}
\overline{D}^{\dot{T}}_{\;\dot{B'}} & 0\end{array} \right) $} .
\end{equation}
Now let us write the eighth invariant term as follows:
\begin{eqnarray}
\label{constructinv8}
&&n_{\nu^{\prime}} (n_{\sigma} \Gamma^{\sigma} \Gamma_{\lambda} 
\Lambda)^{\alpha}{}_{\beta^{\prime}} = 
n_{\nu^{\prime}}n_{\sigma}(\Gamma^{\sigma})^{\alpha}{}_{\rho}
(\Gamma_{\lambda})^{\rho}{}_{\tau}\Lambda^{\tau}{}_{\beta^{\prime}} 
\nonumber\\
& \longrightarrow & -2 
n_{N'{\dot{N}}'}n_{S\dot{S}} \mbox {$ \left( \begin{array}{cc}
0 & \varepsilon^{S}_{\;R} \varepsilon^{\dot{S}\dot{A}}  \\
\varepsilon^{\dot{S}}_{\;\dot{R}}\varepsilon^{SA} & 0 
\end{array} \right) $}\mbox {$ \left( \begin{array}{cc}
0 & \varepsilon_{LT} \varepsilon_{{\dot{L}}}^{\;{\dot{R}}}  \\
\varepsilon_{\dot{L}\dot{T}}\varepsilon_{L}^{\;R}& 0 
\end{array} \right) $} \mbox{$ \left( \begin{array}{cc}
0 & D^{T}_{\;B'} \\
\overline{D}^{\dot{T}}_{\;\dot{B'}} & 0 \end{array} \right) $}.
\end{eqnarray}
The final result for the eighth invariant is
\begin{equation}
\label{inv8}
n_{\nu^{\prime}} (n_{\sigma} \Gamma^{\sigma} 
\Gamma_{\lambda} \Lambda)^{\alpha}{}_{\beta^{\prime}} 
\longrightarrow - 2 n_{N'{\dot{N}}'} \mbox {$ \left( 
\begin{array}{cc}
0 & n_{L}^{\;\dot{A}}\varepsilon_{\dot{L}\dot{T}}D^{T}_{\;B'}  \\
n^{A}_{\;\dot{L}}\varepsilon_{LT}\overline{D}^{\dot{T}}_{\;\dot{B'}} 
& 0 \end{array} \right) $}.
\end{equation}
The ninth invariant structure can be translated in two-spinor form according 
to the following rule:
\begin{eqnarray}
\label{constructinv9}
&&(\Gamma_{\lambda} \Gamma_{\nu^{\prime}} 
\Lambda)^{\alpha}{}_{\beta^{\prime}} = 
(\Gamma_{\lambda})^{\alpha}{}_{\rho}(\Gamma_{\nu'})^{\rho}{}_{\tau}
\Lambda^{\tau}{}_{\beta^{\prime}} = 
(\Gamma_{\lambda})^{\alpha}{}_{\rho}
(\Gamma_{\nu'})^{\rho'}{}_{\tau'}
g_{\rho^{\prime}}{}^{\rho}g^{\tau^{\prime}}{}_{\tau}
\Lambda^{\tau}{}_{\beta^{\prime}} 
\nonumber\\
& \rightarrow & -2 D^{R}_{\;R'}\overline{D}^{\dot{R}}_{\;{\dot{R}}'} 
D_{T}^{\;T'}\overline{D}_{\dot{T}}^{\;{\dot{T}}'}
\mbox {$ \left( \begin{array}{cc}
0 & \varepsilon_{LR} \varepsilon_{\dot{L}}^{\;\dot{A}}  \\
\varepsilon_{\dot{L}\dot{R}}\varepsilon_{L}^{\;A}& 0 \end{array} \right) $}
\mbox {$ \left( \begin{array}{cc}
0 & \varepsilon_{N'T'} \varepsilon_{{\dot{N}}'}^{\;{\dot{R}}'}  \\
\varepsilon_{{\dot{N}}'{\dot{T}}'}\varepsilon_{N'}^{\;R'}& 0 
\end{array} \right) $} 
\mbox{$ \left( \begin{array}{cc}
0 & D^{R}_{\;B'} \\
\overline{D}^{\dot{R}}_{\;\dot{B'}} & 0 \end{array} \right) $}.
\end{eqnarray}
Eventually, after a few algebraic steps with matrices, we get the 
ninth invariant 
\begin{equation}
\label{inv9}
(\Gamma_{\lambda} \Gamma_{\nu^{\prime}} 
\Lambda)^{\alpha}{}_{\beta^{\prime}} \longrightarrow -2 \mbox 
{$ \left( \begin{array}{cc}
0 & \varepsilon_{\dot{L}}^{\;\dot{A}}D^{R}_{\;N'}
\overline{D}^{\dot{R}}_{\;\dot{R'}}D_{T}{}^{T'}
\overline{D}_{\dot{T}{\dot{N}}'}D_{LB'} \\
\varepsilon _{L}^{\;A}\overline{D}^{\dot{R}}_{\;{\dot{N}}'}
D^{R}_{\;R'}\overline{D}_{\dot{T}}^{\;\dot{T'}}D_{TN'}
\overline{D}_{\dot{L}{\dot{B}}'} & 0\end{array} \right) $}.
\end{equation}
Last, but not least, the tenth invariant is built as follows:
\begin{eqnarray}
\label{constructinv10}
&&(n_{\sigma} \Gamma^{\sigma} \Gamma_{\lambda} 
\Gamma_{\nu^{\prime}} \Lambda)^{\alpha}{}_{\beta^{\prime}} = 
n_{\sigma} (\Gamma^{\sigma})^{\alpha}{}_{\rho}
(\Gamma_{\lambda})^{\rho}{}_{\tau}
(\Gamma_{\nu'})_{\chi}{}^{\tau}\Lambda^{\chi}{}_{\beta^{\prime}} 
\nonumber\\
&=& n_{\sigma} (\Gamma^{\sigma})^{\alpha}{}_{\rho}
(\Gamma_{\lambda})^{\rho}{}_{\tau}
(\Gamma_{\nu'})_{\chi'}{}^{\tau'}g_{\tau^{\prime}}{}^{\tau}
g^{\chi^{\prime}}{}_{\chi}\Lambda^{\chi}{}_{\beta^{\prime}} 
\nonumber \\
& \longrightarrow & -2{\sqrt{2}} n_{S\dot{S}} \mbox {$ 
\left( \begin{array}{cc}
0 & \varepsilon^{S}_{\;R} \varepsilon^{\dot{S}\dot{A}}  \\
\varepsilon^{\dot{S}}_{\;\dot{R}}\varepsilon^{SA} & 0 
\end{array} \right) $}\mbox {$ \left( \begin{array}{cc}
0 & \varepsilon_{LT} \varepsilon_{{\dot{L}}}^{\;{\dot{R}}}  \\
\varepsilon_{\dot{L}\dot{T}}\varepsilon_{L}^{\;R}& 0 
\end{array} \right) $}\mbox {$ \left( 
\begin{array}{cc}
0 & \varepsilon_{N'K'} \varepsilon_{{\dot{N}}'}^{\;{\dot{T}}'}  \\
\varepsilon_{{\dot{N}}'{\dot{K}}'}\varepsilon_{N'}^{\;T'}& 0 
\end{array} \right) $}\nonumber\\
&\times& D^{T}_{\;T'}\overline{D}^{\dot{T}}_{\;{\dot{T}}'} 
D_{K}^{\;K'}\overline{D}_{\dot{K}}^{\;{\dot{K}}'}\mbox{$ 
\left( \begin{array}{cc}
0 & D^{K}_{\;B'} \\
\overline{D}^{\dot{K}}_{\;\dot{B'}} & 0 \end{array} \right) $} .
\end{eqnarray}
At the end of the day, when all dust gets settled we obtain the final 
invariant in the form
\begin{equation}
\label{inv10}
(n_{\sigma} \Gamma^{\sigma} \Gamma_{\lambda} \Gamma_{\nu^{\prime}} 
\Lambda)^{\alpha}{}_{\beta^{\prime}} \longrightarrow 
-2{\sqrt{2}} \mbox {$ \left( \begin{array}{cc}
n_{L}^{\;\dot{A}}D^{T}{}_{T'}\overline{D}_{\dot{K}}^{\;\dot{K'}}
\overline{D}^{\dot{K}}_{\;{\dot{B}}'}
{\overline D}_{\dot{L}{\dot{N}}'}D_{KN'} & 0\\
0 & n^{A}_{\;\dot{L}}\overline{D}^{\dot{T}}_{\;\dot{T'}}
D_{K}^{\;K'}D^{K}_{\;B'}D_{LN'}\overline{D}_{\dot{K}{\dot{N}}'} 
\end{array} \right) $} .
\end{equation}

\subsection{The weight functions multiplying the invariants}

A rather tedious but straightforward calculation gives a system of 
$10$ equations for the $10$ coefficient functions $\alpha, ..., \kappa$ 
in (\ref{ansatz}) as found in (See equations (3.6)-(3.15) 
in \cite{anguelova}). It was also found there that one 
can easily express the algebraic solutions for 
$\alpha, \beta, \gamma, \delta, \varepsilon, \theta, \tau, \omega$ 
in terms of the $(\pi, \kappa)$ pair in case of de Sitter space, i.e.
(hereafter we set $n=4$ in the general formulae of \cite{anguelova},
since only in the four-dimensional case the two-component-spinor
formalism can be applied)
\begin{eqnarray}
\label{alg}
\omega &=& \frac{2mC \kappa + ((A+C)^{2}-m^2) \pi}
{(m^{2}+R^{-2})}, \nonumber\\
\theta &=& \frac{((A-C)^{2}-m^{2}) \kappa - 2mC \pi}
{(m^{2}+R^{-2})}, \nonumber\\
\tau &=& \frac{2mC \kappa + ((A+C)^{2}-m^2) \pi}
{(m^{2}+R^{-2})}, \nonumber\\
\varepsilon &=& \frac{-([(A-C)^2 + 2/R^2] +m^2) 
\kappa + 2mC \pi}{(m^{2}+R^{-2})}, \nonumber\\
\alpha &=& - \tau - 4\pi , \nonumber\\
\beta &=& 2 \omega , \nonumber\\
\gamma &=& \varepsilon - 2 \kappa , \nonumber\\ 
\delta &=& 2\varepsilon + 4 (\kappa -\theta) , 
\end{eqnarray}
where we have used the relation $C^2 - A^2 = 1/R^2$. 

Furthermore, from (\ref{alg}) we can immediately see that
\begin{equation}
\label{Sym}
\tau = \omega \qquad {\rm and} \qquad \varepsilon + \theta 
= - 2 \kappa . 
\end{equation}
On using (\ref{Sym}) the differential equations for $\kappa$ and $\pi$, the 
equations (3.14) and (3.15) of \cite{anguelova} acquire the form
\begin{eqnarray} 
\label{kp}
-(A+C) \theta + \kappa^{\prime} + \frac{1}{2} (A-C)  
\kappa + m \pi &=& 0 , \nonumber \\
(C-A) \omega + \pi^{\prime} + \frac{1}{2} (A+C) \pi 
+ m \kappa &=& 0 ,
\end{eqnarray}
where $\theta$ and $\omega$ are given in (\ref{alg}). Clearly one can 
solve algebraically the second equation for $\kappa$. By differentiating 
the result one obtains also $\kappa^{\prime}$ in terms of 
$\pi$, $\pi^{\prime}$ and $\pi^{\prime \prime}$, 
and substitution of these in the 
first equation yields a second order ODE for $\pi (\mu)$.
Now let us look at the system (\ref{kp}) in case of de Sitter 
spacetime. On inserting $A$ and $C$ from
(\ref{AC}) and passing to the globally defined variable $z = \cos^2
\frac{\mu}{2R}$ (see Sec. III), we obtain the following 
differential equation for $\pi$:
\begin{equation}
\label{pisol}
\left[P_{2}\frac{d^2}{dz^2}+ P_{1}\frac{d}{dz}+ P_{0}\right]\pi = 0,
\end{equation}
where $P_{2}$ in (\ref{pisol}) is a quartic polynomial in $z$, i.e.
\begin{equation}
\label{P0} 
P_{2} = 4 \left[m^{2} R^{2}+1 \right] z^4 
-4(2 m^{2} R^{2}+3) z^3
+4(m^{2}R^{2}+2)z^{2}.
\end{equation}
Similarly, $P_{1}$ in (\ref{pisol}) is a cubic polynomial in $z$,
\begin{equation} 
\label{P1}
P_{1} = 16 \left[m^{2}R^{2}+1\right] z^3 
-12 \left[2m^{2}R^{2}+5 \right]z^2 
+ 8 \left(m^{2}R^{2}+2\right) z.
\end{equation}
Last, $P_{0}$ in (\ref{pisol}) is a quadratic polynomial in $z$, i.e.
\begin{equation} 
\label{P2}
P_{0} = \left(4m^{4}-19m^{2}
+32 m^{2}R^{2}+9\right) z^{2} 
- \left(4m^{4}-14m^{2}+32m^{2}R^{2}+21\right)z
-3m^{2}R^{2}-6.
\end{equation}
On making the substitution $\pi (z) = \sqrt{z} \, \tilde{\pi} (z)$, 
(\ref{pisol}) becomes an equation of the type
\begin{eqnarray} 
\label{Heun}
&&z(z-1)(z-a)y^{\prime \prime}(z) + \left\{ (b+c+1)z^2 -
\left[b+c+1+a(d+e)-e\right]z +ad
\right\} y^{\prime} (z) \nonumber \\
&&+ (bc\, z-q) y(z) = 0,
\end{eqnarray}
where the parameters in (\ref{Heun}) take the values
\begin{eqnarray} 
a &=& \frac{(m^{2}R^{2}+2)}{(m^{2}R^{2}+1)}, \nonumber\\
b &=& 2+imR, \nonumber\\
c &=& 2-imR, \nonumber\\
d &=& e = 3, \nonumber\\
q &=& -\frac{(m^{4}R^{4}+7m^{2}R^{2}+10)}{(m^{2}R^{2}+1)}.
\label{(6.40)}
\end{eqnarray}
The equation (\ref{Heun}) is known as Heun's differential equation 
\cite{handbook,heundiff}.
Its solutions, here denoted by ${\rm Heun}(a,b,c,d,e,q;z)$, have
in general four singular points, i.e.
$z_{0}=0,1,a,\infty$. Near each singularity the function behaves as 
a combination of two terms that are powers of $(z-z_0)$ with the 
following exponents:
$\{0, 1-d\}$ for $z_0 = 0$, $\{0, 1-e\}$ for $z_0=1$, $\{0, d+e-b-c\}$ 
for $z_0=a$, and $\{b,c\}$ (that is, 
$z^{-b}$ or $z^{-c}$) for $z\to \infty$.

We now insert into the second of Eq. (6.34) the first of Eq. (6.32),
finding eventually
\begin{equation}
\kappa=f^{-1} \left \{ \left[(A-C)((A+C)^{2}-m^{2})
-{1\over 2}(A+C)(m^{2}+R^{-2})\right] \pi
-(m^{2}+R^{-2})\pi' \right \},
\label{(6.41)}
\end{equation}
where
\begin{equation}
f \equiv m (m^{2}+R^{-2}+2C(C-A)),
\label{(6.42)}
\end{equation}
and $\pi$ and $\pi'$ are meant to be expressed through the Heun function
${\rm Heun}(a,b,c,d,e,q;z)$. Eventually, all weight functions can be
therefore expressed through such Heun function, although the calculational 
details are a bit cumbersome.

\subsection{Peierls bracket for gravitinos}

The expression (6.3) for the gravitino propagator can be written,
concisely, in the form
\begin{equation}
S_{\lambda \nu'}^{\; \alpha \beta'} \longrightarrow
\sum_{k=1}^{10} w_{k} \; 
{ }_{k}S_{L{\dot L}N' {\dot N}'}^{\; A{\dot A} B' {\dot B}'},
\label{(6.43)}
\end{equation}
where, as $k$ ranges from $1$ through $10$, $w_{k}=\alpha,\beta,...,\kappa$
in (6.3), while the 
${ }_{k}S_{L{\dot L}N'{\dot N}'}^{\; A{\dot A}B'{\dot B}'}$
are the $10$ spinor invariants written down in subsection 6.1.  
Two further indices are needed to characterize each $w_{k}$
function, i.e. $j$ which labels the four singular points at $z=0,1,a,\infty$
and the subscript $F$ to denote the Feynman prescription to approach such 
singular points, i.e. from the above along the positive real axis.
Thus, the definition of Peierls bracket that we propose bears analogies
with Eqs. (5.29) and (5.30), with $\psi_{\nabla}$ 
and $\chi_{\nabla}$ obtained from the covariant derivative 
of the Rarita--Schwinger potential (see appendix), while  
\begin{equation}
({\rm Re}G_{F}) \longrightarrow
{\rm Re}\left(\sum_{k=1}^{10}w_{k}^{(j_{F})} \;
{ }_{k}S_{L{\dot L}N'{\dot N}'}^{\; A{\dot A}B'{\dot B}'}\right).
\label{(6.44)}
\end{equation}

\section{Concluding remarks}

Our paper has been devoted to geometric constructions of current interest 
in theoretical physics. Its original contributions, of structural 
nature, are as follows:
\vskip 0.3cm
\noindent
(i) A two-component-spinor analysis of geometric invariants contributing
to the gravitino propagator in four-dimensional de Sitter spacetime.
\vskip 0.3cm
\noindent
(ii) A Peierls bracket for massive spin-1/2 and spin-3/2 fields in
de Sitter spacetime has been proposed, by relying upon the same tools
as in item (i) above.

Our use of positive- and negative-frequency Green functions to 
re-express the Peierls bracket is also of some interest, by virtue of
the more direct link with the Feynman Green function.
It now remains to be seen whether our brackets can be exploited to
study quantum field theories in de Sitter spacetime from a
modern perspective. At a technical level, it would be also interesting 
to exploit the work in Ref. \cite{gaspar} to re-express all weight
functions in subsection 6.B through hypergeometric functions.  

\appendix
\section{Rarita--Schwinger potentials}

The gravitinos of supergravity are described by spinor-valued one-forms
$\psi_{\mu}^{A}$, where $\mu$ is the Greek index used to denote the
one-form nature. Bearing in mind that the soldering form is obtained by
contracting the tetrad $e_{a}^{\; {\hat c}}$ with the 
Infeld-van der Waerden symbols $\tau_{{\hat c}}^{\; B{\dot B}}$
according to
\begin{equation}
e_{a}^{\; B{\dot B}}=e_{a}^{\; {\hat c}} \; 
\tau_{{\hat c}}^{\; B{\dot B}},
\label{(A1)}
\end{equation}
one can write the spatial components of the gravitino in the form
\begin{equation}
\psi_{A \; i}=\Gamma_{\; AB}^{{\dot C}} \;
e_{\; {\dot C}i}^{B},
\label{(A2)}
\end{equation}
where $\Gamma$, the Rarita--Schwinger potential, 
can be obtained from a spinor field $\alpha$ according to
\cite{aichelburg}
\begin{equation}
\Gamma_{\; B{\dot B}}^{A}=\nabla_{B{\dot B}} \; \alpha^{A}.
\label{(A3)}
\end{equation}
It obeys the equations
\cite{GRQC9911051} ($\Lambda$ being the cosmological constant,
and $\Phi$ being the trace-free part of Ricci)
\begin{equation}
\varepsilon^{BC}\nabla_{{\dot A}(A} \;
\Gamma_{\; B)C}^{{\dot A}}=-3 \Lambda \alpha_{A},
\label{(A4)}
\end{equation}
\begin{equation}
\nabla^{B({\dot B}} \; \Gamma_{\; \; BC}^{{\dot A})}
={\overline \Phi}_{\; \; \; \; \; \; \; C}^{{\dot A}{\dot B}L}
\; \alpha_{L},
\label{(A5)}
\end{equation}
and the gauge-transformation law
\begin{equation}
{\widehat \Gamma}_{\; \; BC}^{{\dot A}}
={\Gamma}_{\; \; BC}^{{\dot A}}
+\nabla_{\; B}^{{\dot A}} \; \nu_{C}.
\label{(A6)}
\end{equation}

In the Peierls bracket proposed in Sec. VI.C, the role of
$\psi_{\nabla}$ and $\chi_{\nabla}$ in (5.30) will be played 
by covariant derivatives of such spinor-valued one-forms, or, in 
purely two-component-spinor language, by spinor covariant derivatives
of the Rarita--Schwinger potential occurring in (A2)--(A6).
A part of the existing literature on supergravity prefers instead to
omit spinor indices, writing simply $\psi_{\mu}$ for gravitinos.
With this notation, one can say that, to the functional derivative
$A_{,i}$ in the definition (2.11) there corresponds the covariant
derivative \cite{freedman}
\begin{equation}
D_{\nu}\psi_{\rho}(x)=\partial_{\nu}\psi_{\rho}(x)
-\Gamma_{\; \nu \rho}^{\sigma} \; \psi_{\sigma}(x)
+{1\over 2}\omega_{\nu ab}\sigma^{ab}\psi_{\rho}(x),
\label{(A7)}
\end{equation}
where $\Gamma_{\; \nu \rho}^{\sigma}$ are the Christoffel symbols,
$\omega_{\nu ab}$ is the spin-connection, and $\sigma_{ab}$ is
proportional to the commutator of ``flat'' $\gamma$-matrices, i.e.
\begin{equation}
\sigma_{ab}={1\over 4}[\gamma_{a},\gamma_{b}].
\label{(A8)}
\end{equation}

\acknowledgments

The authors are grateful to the Dipartimento di Scienze Fisiche
of Federico II University, Naples, for hospitality and support.
Conversations with Ebrahim Karimi are gratefully acknowledged.
One of us, G.E., dedicates the present work to Maria Gabriella,
Maria Giuseppina and Gennaro.

\vskip 0.3cm
\noindent
{\bf Note added in proof}: In a forthcoming paper in General Relativity
and Gravitation by us (arXiv:0907.3634 [hep-th]) we have extended the 
analysis of the gravitino propagator in four-dimensional de Sitter space
to the classification of the $z$ region, in the sense that we find two
ranges of values of $z$, in which the weight functions can be divided into
dominant and subdominant families.
\end{document}